# Photothermal gas detection using a miniaturized fiber Fabry-Perot cavity

KAROL KRZEMPEK,[1,*] PIOTR JAWORSKI,[1] LUKAS TENBRAKE,[2] FLORIAN GIEFER,[2] DIETER MESCHEDE,[2] SEBASTIAN HOFFERBERTH,[2] HANNES PFEIFER[2]

[1]*Laser Spectroscopy Group, Department of Electronics, Photonics and Microsystems, Wroclaw University of Science and Technology, Wroclaw, Poland*

[2]*Institute of Applied Physics, University of Bonn*

**karol.krzempek@pwr.edu.pl*

**Abstract:** We demonstrate a robust and miniaturized fiber Fabry-Perot cavity-based sensor for photothermal spectroscopic signal retrieval. The proof-of-concept experiment involved the use of a near-infrared *pump* laser to detect methane molecules on an isolated overtone 2ν3 R(4) transition located at 6057.1 $cm^{-1}$. The photothermal-related modulation of the gas refractive index was induced at the center of the interferometer, which was filled with the sample. Subsequently, the phase change of the resonating probe beam was measured as a shift in the reflected beam intensity, which was proportional to the methane concentration. A normalized noise equivalent absorption coefficient of $7.06\times10^{-8}$ $cm^{-1}$ W $Hz^{-1/2}$ was achieved, suggesting significant potential for the design of small and versatile gas detectors with excellent detectivity. We discuss future improvements of the proposed photothermal gas detection approach.

## 1. Introduction

Accurate, selective and in vivo monitoring of gas particles is critical in industry, environmental protection, and medical applications [1]. Laser spectroscopy is a commonly used gas monitoring technique, which offers low detection limits, long-term stability, real-time monitoring and non-intrusive measurements [2]. However, most laser-based spectroscopic techniques rely on the direct-absorption approach, where the properties of the beam interacting with the gas sample are directly probed to extract the analyte composition and concentration. Therefore, taking advantage of strong transitions of gas molecules located in the mid-infrared (mid-IR) necessitates the integration of low-noise photodetectors designed to operate in this wavelength region, e.g. mercury cadmium telluride (MCT) detectors. Constraints stemming from usage of mid-IR detectors motivated researchers to develop sensor configurations having sub-ppm detection capabilities, while relying on well-established and cost-effective components and devices working in the near-infrared (near-IR) wavelength region. However, not all gas molecules have strong fingerprint spectra in the range from 1 μm to 2 μm. As a result, various techniques for gas detection have been developed that combine the advantages of using mid-IR lasers as the excitation sources with the simplicity and low cost of near-IR detectors and optics. Such sensors are based on an indirect detection principle.

The photothermal (PT) effect is an excellent example of a physical effect that can be exploited to build a sensitive and selective gas detector that relies on a non-direct spectroscopic technique [3]. With the first experimental demonstration in the 1970s, photothermal spectroscopy (PTS) of gases is currently experiencing a renaissance. This development is based on advances in optics manufacturing, low-noise near-IR detectors and novel approaches to spectroscopic signal retrieval combined with efficient denoising and signal processing techniques.

In PTS, a *pump* laser is used to excite the gas molecules. The wavelength emission of the laser must match the absorption profile of the analyte, similar to traditional direct-absorption techniques. The energy absorbed by the particles is released to its environment mostly via nonradiative collision energy transfer, resulting in a local temperature rise and gradient of the gas sample along the *pump* laser's pathlength. Temperature gradients cause local variations in gas density, which manifest themselves as changes in the gas refractive index (RI). The majority of published gas sensors relying on the PTS effect probe the RI modulations using an additional laser known as the *probe*. The separation of the *pump* and *probe* sections of the setup is the distinguishing characteristic of non-direct detection techniques, which enables optimizing each component separately for its particular task. For example, the *probe* section of the sensor can be conveniently built based on inexpensive and widely available near-IR lasers, detectors and optics. The amplitude of the RI modulation induced in the gas sample via the PT effect is minuscule, however. Therefore, different methods for retrieving the spectroscopic signal have been developed. Signal retrieval based on Fabry-Perot (FPI) and Mach-Zehnder (MZI) interferometers is in widespread use as it offers the required RI modulation readout sensitivity. Most of the recently published PTS sensor configurations take advantage of the inexpensive and reliable fiber components designed for 1.55 μm wavelengths. Yet, the indirect detection schemes of the PTS technique are rarely documented, as most sensor configurations use *pump* and *probe* lasers that emit in the near-infrared (near-IR) and fibers with hollow cores [4,5]. Sensor configurations capable of targeting mid-IR gas transitions in MZI-based configurations have also been demonstrated, however they suffer from increased footprint and complexity, as additional beam combining optics is a requirement [6–8]. Moreover, MZI-based PTS signal extraction in most cases necessitates active stabilization of the interferometer at the quadrature point or acousto-optical frequency shifters combined with signal demodulation at MHz-level carrier frequencies.

The alternative approach uses the high sensitivity of the FPIs for precise RI modulation probing. Similarly to MZI-based detectors, most publications present results in which high sensitivity and long-term stability are a consequence of an all-in-fiber-based configuration of the sensors, designed for operation with near-IR lasers only, thus severely limiting their versatility [9–13]. Recently, FPI-based PTS sensors capable of probing gases in the mid-IR were also presented [14,15]. However, the low finesse of the interferometer cavities and the necessity of using hollow-core fibers (HCF) or antiresonant hollow-core fibers (ARHCF) with relatively narrow transmission bandwidth as gas absorption cells are still their main limiting factor.

In this article, we demonstrate a novel PTS sensor configuration with a home-made fiber Fabry-Perot cavity (FFPC) used for probing the RI modulation induced in gas samples by the PT effect. As outlined and reviewed in [16,17], the FFPCs are miniaturized, they can be fabricated from commonly available materials and are characterized by high finesse, passive stability and robustness: excellent conditions when designing PTS sensors. Our proof-of-concept experiment involved the detection of trace concentrations of methane ($CH_4$) by exciting gas molecules directly inside the FFPC. The PT-induced RI modulation of the gas inside the FFPC is translated into variations of the optical power reflected from the cavity, which can be converted to an electrical signal and analyzed in analogy to conventional direct absorption gas spectrometers. The experimental results prove, that the FFPCs can serve as an excellent tool for probing the PT-induced gas RI modulation. Moreover, the proposed configuration does not limit the *pump* wavelength, which is used to excite the gas particles, therefore a significant advantage over the HCF- and ARHCF-based PTS sensors is established. We believe that the combination of PTS and FFPC paves the way for the development of novel, highly selective and sensitive gas sensors with a miniaturized footprint.

## 2. Materials and Methods

### 2.1 Miniaturized Fiber Fabry-Perot Interferometer

Most gas molecules of interest exhibit only weak transitions in the near-IR, where FFPCs offer proven high finesse on the order of $10^5$. By exploiting their susceptibility to changes of the refractive index of the intra-cavity medium they can, however, be used as the core part of a PTS sensor. Employing such a custom-developed miniaturized FFPC strongly reduces the footprint of the RI sensor part, eases integration, and makes use of its high level of passive stability. A schematic of the passively-stable FFPC design [17], a size comparison and its implementation in the gas probing chamber is shown in Fig. 1.

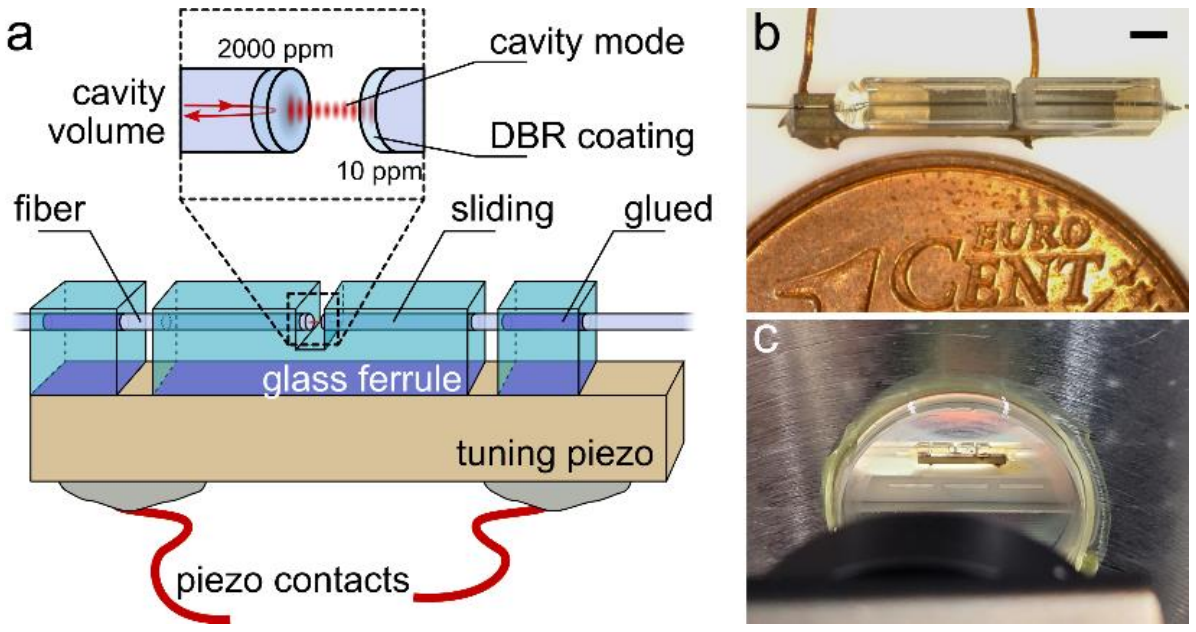

Figure 1. (a) - schematic of the employed FFPC. The alignment and stability are provided by the guiding bore in the slotted glass ferrule. By fixing the fiber mirrors in separate ferrule parts the tunability by the attached piezo is achieved [17]. The inset shows a close-up of the single-sided cavity, where the coupling through the in- and outcoupling mirror dominates the cavity dynamics. (b) - typical size comparison of an FFPC as also used in [18]. Below is a 1 eurocent coin. The black bar above indicates 1 mm. (c) - the integration of the FFPC setup in the employed gas chamber.

The multilayer coating of the FFPC's fiber mirrors was designed for high reflection in the range from 750 nm to 800 nm. We use a single-sided cavity geometry, where the in- and out-coupling mirror has a larger transmission (2000 ppm) than the opposing mirror (10 ppm transmission) and where all measurements are performed in reflection mode of the interferometer. The base finesse at 780 nm probe wavelength was 1370 ± 70 with a coupling depth of ~ 30%. The distance between the fiber mirrors was chosen to be ($d$ = 106 ± 5) µm in order to allow for focusing of the *pump* laser at the center of the interferometer without clipping the beam on the fiber mirrors. The distance can be further extended, however this would not benefit the sensitivity of the PTS sensor as it would affect the alignment and Finesse of the cavity. The waist radius of the fundamental probe resonance in the center of the device is (4.2 ± 0.4) µm with a radius of curvature of the laser-ablation machined fiber mirror surface of ~ 150 µm. A complete discussion of the device design can be found in [17]. Interferometers with finesse up to several $10^5$ can be realized with the same geometry design.

### *2.2* Gas sensor setup

The gas sensor schematic is depicted in Fig. 2. We used a custom-built aluminum gas chamber with total dimensions of 100 x 125 x 27 mm³ to enclose the FFPC and assess the sensor performance under stable conditions. Although further miniaturization of the enclosure gives the possibility of achieving a more compact and versatile sensor, it was not the primary objective in the initial experimental phase presented here. The gas chamber featured four 1/8 inch Swagelok™ gas ports: two were utilized to improve gas exchange within the chamber via constant flow, while the other two were combined with Teflon ferrules, produced as outlined in [19], to ensure an airtight seal of the FFPC fiber feed-through. To minimize back reflection-related noise in the sensor, the fiber of the low-transmission mirror was cut at an 8° angle, as the beam exiting the FFPC in this direction was not used in this experiment. In the proof-of-concept experiment we used a custom second-harmonic generation (SHG) source emitting at 780 nm to probe the FFPC reflection signal (subsequently called the *probe* laser).

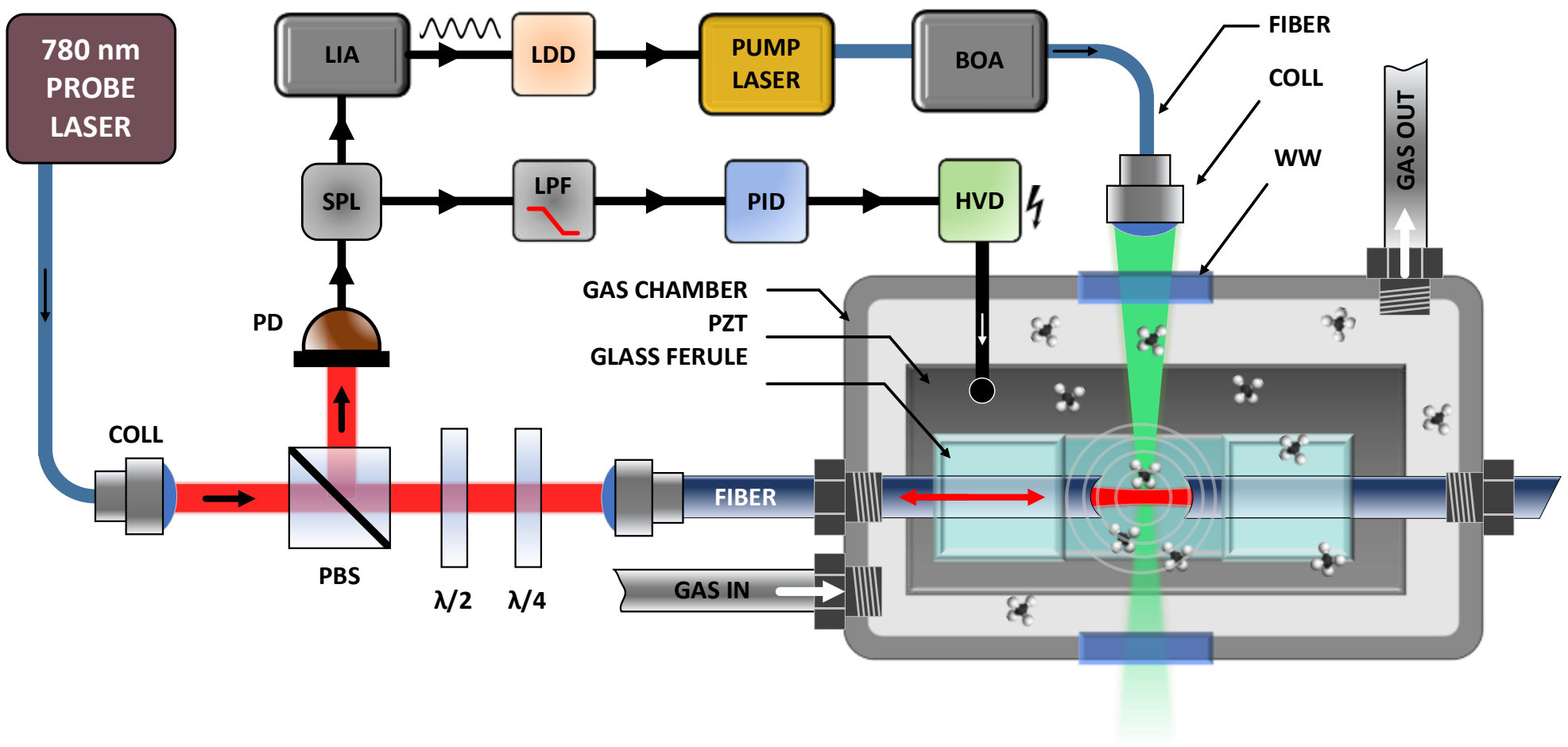

Figure 2. Schematic of the sensor setup. COLL – collimator; LIA – lock-in amplifier; SPL – RF signal splitter; PD – photodiode; PBS – polarization beam splitter; LDD – laser diode driver; LPF – lowpass RF filter; λ/2, λ/4 – waveplates; PID – PID controller; PZT – piezo-ceramic transducer; BOA – semiconductor booster optical amplifier; HVD – high voltage driver; WW – wedged N-BK7 window.

The emission of a narrow-linewidth 1560 nm distributed feedback (DFB, EP1560-0-NLW-B26-100FM, Eblana) laser was amplified in an all-fiber amplifier and focused in a periodically polled lithium niobate crystal (PPLN, MSHG1550-1.0-40, Covesion). The crystal was temperature stabilized at 91.1 °C to achieve quasi-phase matching conditions and efficient generation of the SHG beam. The fiber-based part of the *probe* laser was built using polarization-maintaining fibers and components, ensuring minimum drift in the amplitude of the beam generated via the SHG process.

Ultimately, we aim to develop an FFPC designed for telecom wavelengths, which will significantly reduce the complexity of the proposed PTS sensing scheme. The SHG beam was directed onto a polarization beam splitter (PBS). The transmitted beam was coupled via a collimator to the fiber, leading to the high transmission (2000 ppm) in- and output mirror of the FFPC. The beam reflected from the FFPC returns through the same fiber and is directed by the PBS onto a detector (PDA8A2, Thorlabs), which measures the intensity. A N-BK7 plano-convex lens with f = 50 mm was used along with a bandpass filter (FBH780-10, Thorlabs) before the detector to maximize the signal and decouple the background optical signals. Half- and quarter-waveplates were used to achieve polarization-mode matching to one of the FFPC polarization modes and to optimize the reflection signal amplitude coupled onto the detector. The reflection voltage signal was split and coupled to a proportional-integral-derivative (PID) controller (D2-125, Vescent) and a lock-in amplifier (LIA, model UHFLI, Zurich Instruments). The error signal from the PID controller was fed to a piezoelectric controller (MDT693B, Thorlabs), which controlled the distance between the FFPC mirrors via the integrated piezoceramic element. This ensured a stable lock at a chosen point of the reflection slope (see Fig. 3). A low-pass filter (cutoff frequency 300 Hz) was added to minimize the influence of the PID feedback loop on the spectroscopic signal encoded in the reflection signal (analyzed at $2\times f_0 \sim 2.3$ kHz, explained in Section 3.2). A standard DFB laser diode was used as the *pump* source that excited the $CH_4$ molecules at their overtone transition located at 1651 nm (EP1651-0-DM-B01-FA, Eblana). Its output power was increased to 66 mW in a fiber-pigtailed booster optical amplifier (BOA, model BOA1084P, Thorlabs), which enabled continuous tuning of the output amplitude, if required by the sensor operation. The beam of the *pump* laser was out-coupled from the fiber using a collimator (F810APC-1550, Thorlabs) and focused to a ~ 17 μm spot using a N-BK7 plano-convex lens with 50 mm focal length. The beam was aligned in the center of the FFPC by coupling it through a 1-inch N-BK7 wedge glued to the side of the gas chamber. During the sensor alignment phase red light was coupled to the fiber delivering the *pump* beam, which was used for rough positioning of the coupling optics (fine-tuning was achieved by observing the amplitude of the spectroscopic signal). The unabsorbed light exited the chamber on its opposite side through a second wedge, minimizing reflections and allowing non-complex alignment of the *pump* beam in the optimum spot. To simplify the spectroscopic signal processing and to achieve a higher signal-to-noise ratio (SNR), the wavelength modulation spectroscopy (WMS) measurement technique was implemented in the sensor [20]. This requires modulating the *pump* wavelength with a sinusoidal waveform at an optimized frequency, $f_0$. This was achieved by coupling an appropriate voltage signal from the LIA to the modulation input of the laser driver (LDD, model CLD1015, Thorlabs), in which the DFB laser was mounted. This results in altering the current supplied to the laser and, as a result, a modulation of the emission wavelength. The BOA was working in the saturated regime, thus the *pump* laser amplitude modulation resulting from current variations was negligible. An additional 50 mHz sawtooth ramp was added to the sinewave signal in the LIA when full 2f scans of the $CH_4$ absorption were recorded. The LIA was configured to demodulate the electric signal from the reflection detector at $2\times f_0$, as commonly used in gas sensors employing the WMS signal processing technique. Bulk-optical-based parts used in the experiment were antireflection coated for the respective wavelengths. The gas preparation and delivery infrastructure consisted of gas cylinders with $N_2$ (6.0 purity), certified mixtures of $CH_4$ in $N_2$ (from Air Liquide), a commercial gas mixer (Model 4020, Environics) and a flow meter (GE50, MKS). Necessary connections were made using stainless steel tubes, connectors and valves from Swagelok™.

## 3. Experimental Section

### 3.1 Photothermal signal generation and sensor operation principle

The absorption of light in gas samples gives rise to PT and photoacoustic (PA) effects, which are caused by transitions of the molecules from the ground level to an excited state. The absorbed energy is then released via radiative or non-radiative pathways, with the latter dominating in the near- and mid-IR spectral region. The typical relaxation time of the molecules is $\tau \approx 10^{-6} - 10^{-9}$ s [21], which is mainly the result of collision-based energy transfer to the surrounding molecules. This process converts the optical energy of the laser absorbed by the gas into kinetic energy (i.e. heat) along the pathlength of the propagating beam. If the laser wavelength or amplitude is modulated, a corresponding periodic modulation of PT and PA effects is observed. However, this is only true for small values of laser beam absorption (below saturation) and if the laser is modulated at frequencies that allow the molecules to release the stored energy via the non-radiative pathway ($f \ll \tau^{-1}$). If these requirements are satisfied, the induced periodic perturbation of the thermodynamic equilibrium will result in localized gas expansion, that gives rise to an acoustic wave and simultaneously modulates the RI of the gas sample. In photoacoustic spectroscopy (PAS), the acoustic wave is commonly quantified using microphones or cantilevers (often in combination with acoustic resonators) and quartz tuning forks [22–31]. In contrast, a gas sensor that relies on the PT effect usually takes advantage of an interferometric optical setup, which enhances the sensitivity for measuring the induced the RI modulation. For PAS and PTS, the amplitude of the photo-induced signal can be expressed as [32]:

$$S \propto \frac{\alpha P}{V f} \tag{1}$$

where $\alpha$ is the absorption coefficient of the gas sample, $P$ is the optical power of the light source that excites the molecules, $V$ is the interaction volume. The RI modulation ($\Delta n$) resulting from the PT effect can be expressed as a function of temperature variation ($\Delta T$), based on the Clausius-Mossotti approximation [33]:

$$\Delta n = -\frac{n-1}{T_{abs}}\Delta T \quad (2)$$

where $n$ is the unexcited gas RI and $T_{abs}$ is the temperature of the gas sample. By exciting the gas molecules (and modulating its RI), we induce a phase change in the laser beam propagating through it. In this sensor, the gas sample under test is excited inside the FFPC, and the resulting phase change ($\Delta\varphi$) is given by [32]:

$$\Delta\varphi = -\frac{2\pi L}{\lambda}\frac{(n-1)}{T_{abs}}\Delta T \quad (3)$$

where $L$ is the interaction length, $\lambda$ is the wavelength of the *probe* laser beam. Here, the *probe* laser is coupled into the FFPC ($d$ ~ 106 μm) and the reflected beam intensity ($I_R$) is measured to extract the RI modulation. The sensor described in this article takes advantage of the high finesse of the interferometer ($F$ ~1370 ±70), which results in a significant enhancement of the $\Delta\varphi$ readout sensitivity. The reflection characteristic of the FFPC used in this experiment was measured by scanning the distance between the mirrors via a voltage applied to the PZT. The result is plotted in Fig. 3.

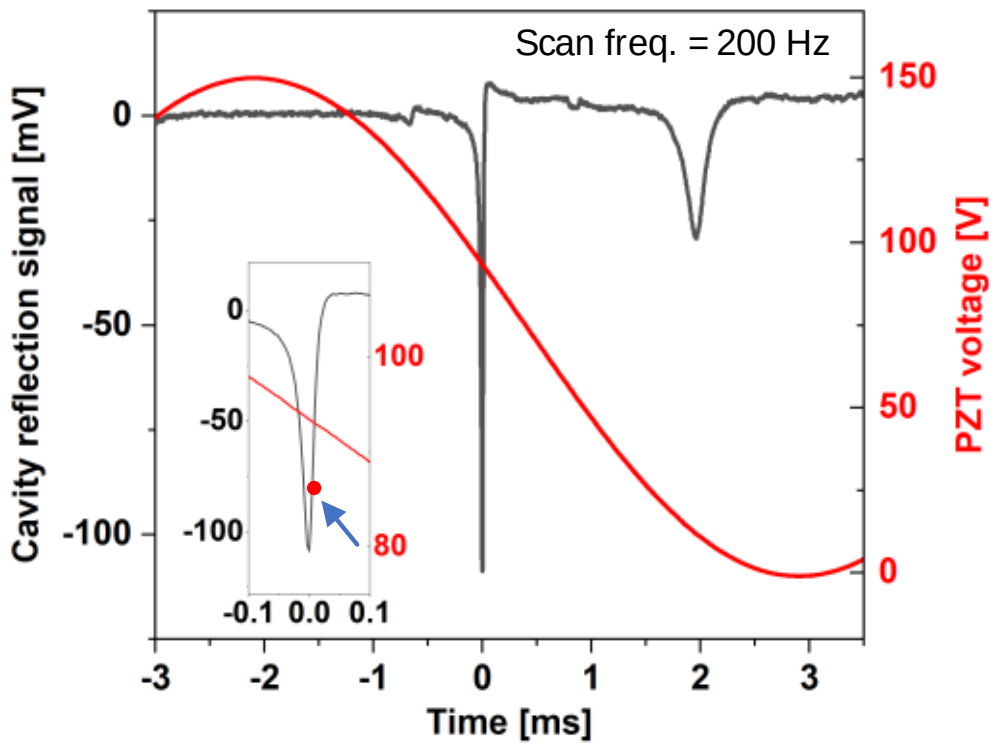

Figure 3. Reflection signal from the cavity measured with the 780 nm *probe* laser by applying a 200 Hz, 150 V sinewave signal to the PZT used for altering the distance between the FFPC mirrors. The inset shows a closeup of the peak with the PID lock-point highlighted with a red spot.

In the 150 V scanning range, the FFPC exhibits two separate reflection dips. The deep, narrow dip is the typical fundamental mode of the FFPC. Due to the mode filtering of the back-reflected light and light from the resonator its shape corresponds to a dispersion altered Lorentzian function [34]. The shallower dip is not used in the experiment and corresponds to a higher order mode of the resonator. If we assume that the FFPC is in a steady state (fixed $I_0$, $d$, $\lambda_0$, $F$), the only parameter that modulates the $I_R$ is the RI of the gas trapped in the mode profile of the cavity, which is varied by the auxiliary *pump* laser via the PT effect. The wavelength of the *pump* laser is additionally modulated with a sinewave signal and the resulting periodic change in $I_R$ is monitored by the detector and fed to the LIA. As we have implemented the WMS spectroscopic signal extraction technique, any parasitic signals coupled to the reflection amplitude (e.g., acoustic and mechanical vibrations or electronic noise) having frequencies beyond the narrow demodulation bandwidth of the LIA will be rejected, significantly improving the detectivity of the sensor. Nevertheless, active stabilization of the FFPC at the optimum point of the reflection slope is crucial for achieving high SNR and stable long-term sensor operation in constant gas flow conditions. During the initial setup of the sensor a constant voltage of 75 V is applied to the PZT (to ensure bi-directional corrections with the PID error signals can be achieved) and the wavelength of the *probe* laser is tuned manually until the optimum spot is reached on the steeper slope of the reflection dip (highlighted with a red dot in Fig. 3). In this setup a maximum 2f PTS signal amplitude was observed at a lock point corresponding to a value of ~25% of the peak-to-peak reflection, similar to the work [35]. At this point, the PID controller is enabled, which allows long-term operation of the sensor under varying conditions.

### 3.2 Sensor optimization

Achieving a high sensor sensitivity requires proper optimization of several parameters. Firstly, an appropriate absorption line has to be selected. In the proof-of-concept experiment a near-IR absorption band of $CH_4$ was chosen, which has negligible interference from other atmospheric gases. A simulation based on the HITRAN database [36] is presented in Fig. 4. The *pump* laser allowed targeting an isolated overtone 2ν3 R(4) transition located at 6057.1 $cm^{-1}$.

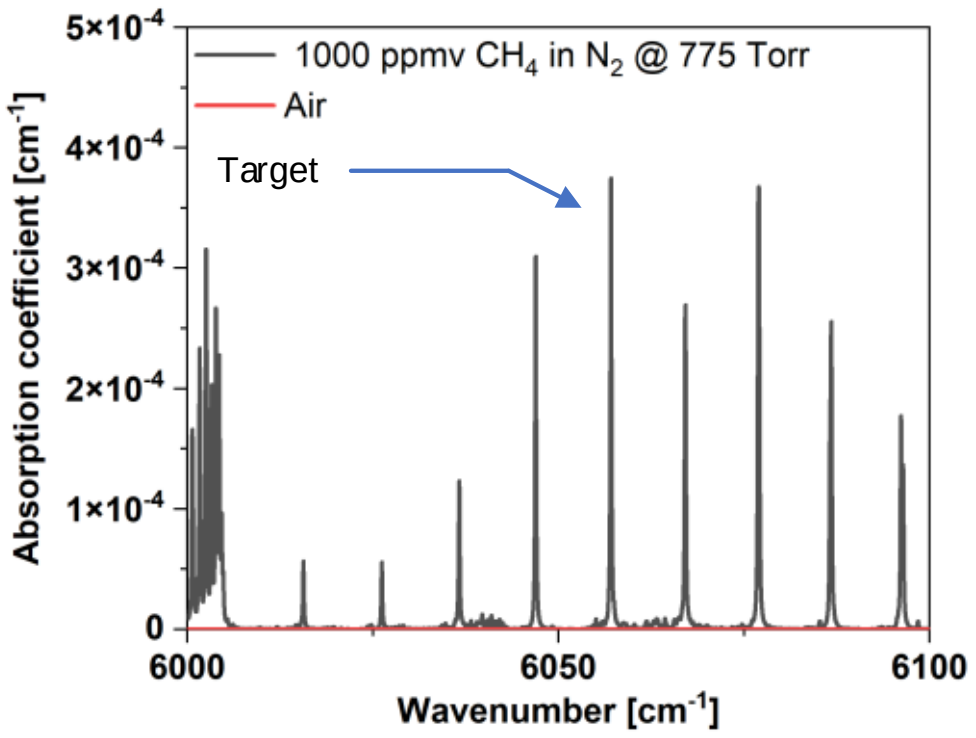

Figure 4. Near-IR absorption spectrum of $CH_4$ with the target transition indicated by an arrow. The simulation parameters are listed in the graph.

The following optimization steps were performed with the PID feedback loop locked to the optimum cavity reflection signal. Gas samples were flowing through the gas chamber at a constant rate of 100 sccm. The second parameter that requires optimization is the sinewave modulation frequency ($f_0$), which is used to modulate the wavelength of the *pump* laser. The process is less straightforward, when compared to classical direct-absorption sensors employing the WMS signal processing technique. In the case of the developed sensor three variables have to be considered. The collision-based energy transfer between the excited molecules is the main contributor to the heat transfer in the gas sample, however, the rate at which this process occurs is limited, as described in the Introduction. Therefore, a decay in the amplitude of the PTS signal is observed with increasing modulation frequency, according to Equation (1). Employing Hertz-level $f_0$ is thus tempting, but results in a significant increase in background noise picked up by the sensor, due to the 1/f characteristic. Moreover, if interferometers that require active stabilization are used to extract the PTS signal, a proper feedback loop cutoff frequency must be included in the considerations. In conclusion, the optimal value must be determined experimentally, for each sensor setup individually, using the approach described in [14]. The process is split into two steps, which consist of independent measurements of the noise density (in pure $N_2$) and 2f PTS signal amplitude (using 1000 ppmv of $CH_4$ diluted in $N_2$), as a function of the *pump* laser sinewave modulation frequency. In both cases the $f_0$ is swept from 100 Hz to 100 kHz (500 points were registered). This process is straightforward when using the UHFLI LIA, as it allows the use of the built-in oscillators to generate the sweep and simultaneously demodulate the electric signal at the given frequency, with optimized filter settings. By dividing the 2f PTS signal values by the FFPC noise the SNR characteristic is obtained, which clearly indicates the frequency at which the sensor will achieve optimal performance. The measurement results are plotted in Fig. 5. Based on experimental verification, the highest SNR in this sensor configuration was achieved for $f_0$ = 1150 Hz, and this modulation frequency was used in the following experiments and performance evaluation. The third parameter requiring optimization is the modulation depth, which is correlated with the linewidth of the targeted gas transition. During the measurement, the *pump* laser was tuned to the center of the selected $CH_4$ absorption line and the 2f PTS signal amplitude was measured as a function of the *pump* laser wavelength modulation depth (from 3 GHz to 45 GHz with 500 measurement points).

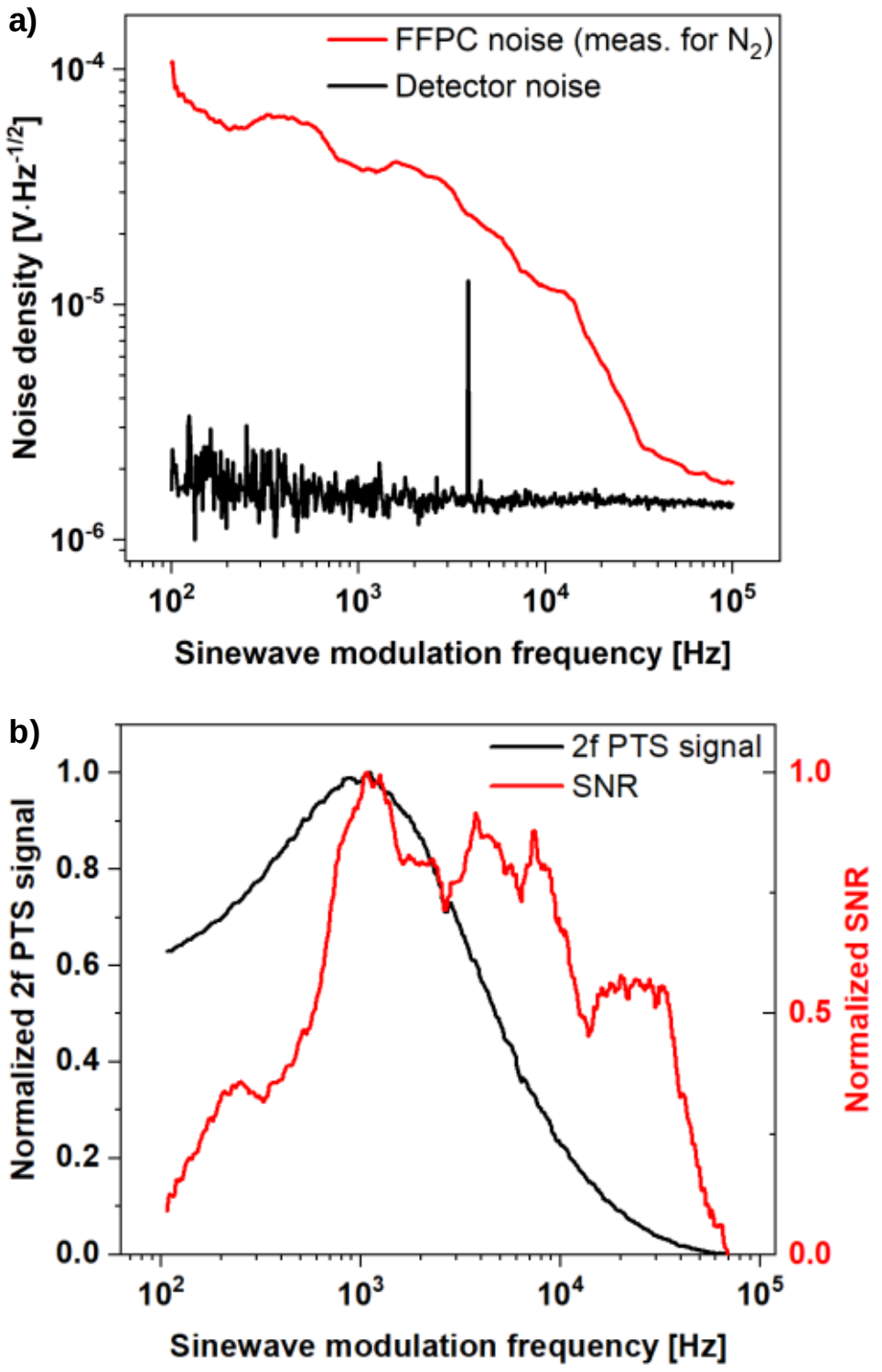

Figure 5. *Pump* laser sinewave modulation frequency optimization. (a) The red line presents the noise density spectrum of the locked FFPC (with 300 Hz locking bandwidth) plotted as a function of the sinewave modulation frequency. The black plot shows the noise density spectrum of the detector (light blocked). b) The black plot presents the normalized 2f PTS signal as a function of the sinewave modulation. The red plot shows the normalized SNR, which was calculated by performing a point-by-point division of the 2f PTS signal and the FFPC noise. The 2f PTS signal amplitude was registered for 1000 ppmv $CH_4$ in $N_2$ at 775 Torr pressure.

The results are plotted in Fig. 6a. Fig. 6b shows the full 2f spectra registered for four chosen values of the modulation depth with a 20 s long sweep of the *pump* center wavelength. The highest 2f PTS signal was observed for a *pump* wavelength modulation depth equal to 10.7 GHz, which is comparable to the values achieved in other sensors targeting this $CH_4$ transition in the near-IR at ambient pressure values [37]. The full-width at half-maximum (FWHM) of the $CH_4$ transition at 775 Torr is 5.1 GHz, therefore, the experimentally determined value of modulation depth vs. FWHM is close to the theoretical value of m = 2.2 [38]. Finally, the sensor response to a varying *pump* power was examined. Equation 1 that describes the amplitude of the PT-generated signal takes into account the optical power, which is used to excite the gas molecules. In principle, the amplitude of the observed RI modulation is proportional to the *pump* power. However, parasitic absorption of the beam in the optical components of the sensor (windows, lenses, filters, etc.) can result in background noise, which deteriorates the SNR. Therefore, a proper assessment of the sensor response to varying *pump* power is required in the tested configuration. For this evaluation, we exploited the functionality of the BOA, which was used to boost the optical power of the DFB laser. By varying the DC current delivered to the component, we were able to control its gain and thus the optical power exciting the $CH_4$ molecules in the FFPC.

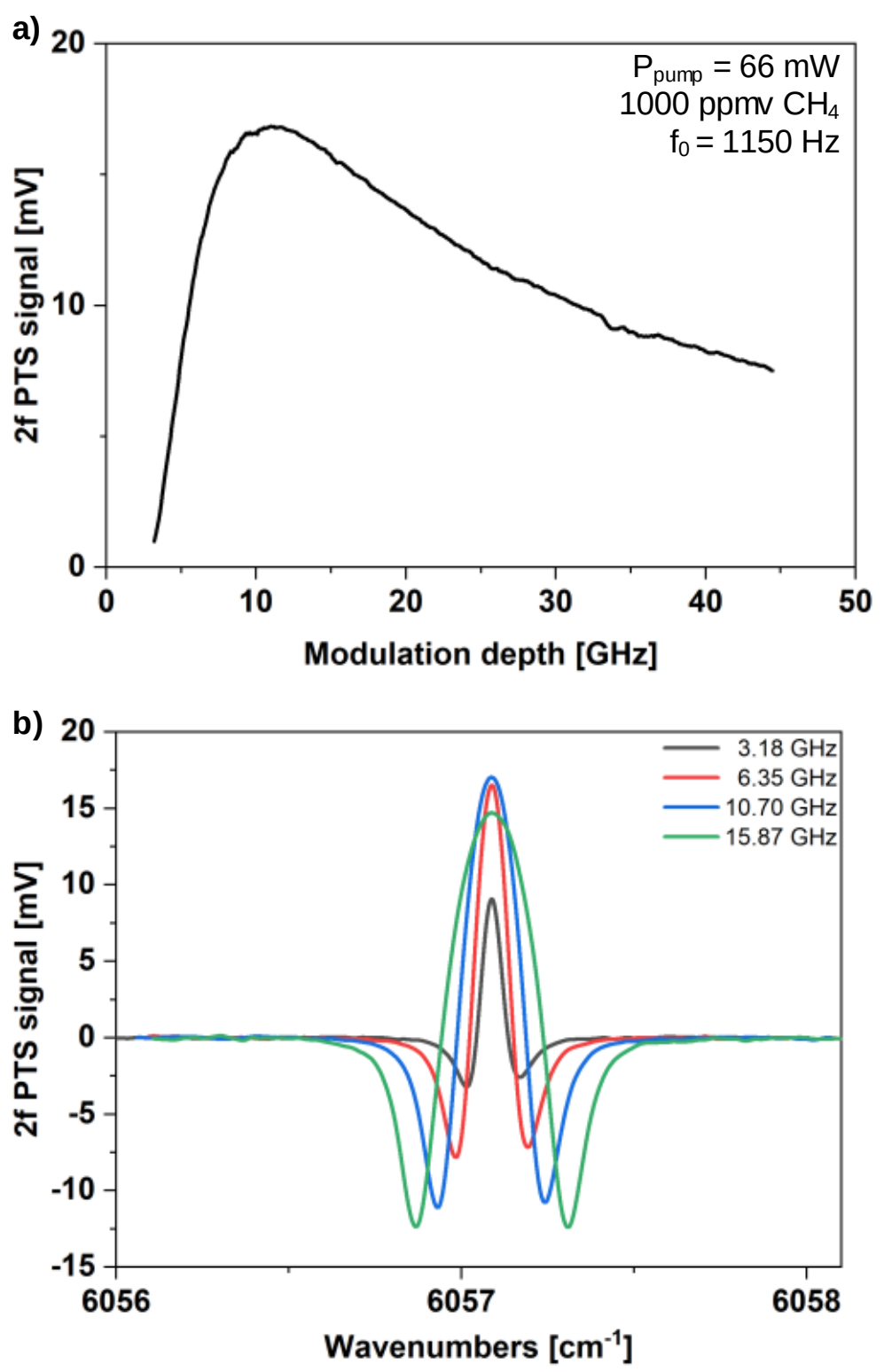

Figure 6. (a) Sensor response for varying *pump* sinewave modulation depth. (b) Full 2f PTS signals registered for modulation depth values as listed in the legend. The measurement conditions are listed in the graphs.

During the experiment, the *pump* laser wavelength was tuned to the center of the targeted gas transition and the maximum 2f PTS signal amplitude was registered for 20 setpoints of the excitation power. The highest output power of the BOA was 66 mW at 600 mA of current (measured after transmission through the wedge window mounted in the gas chamber). Next, the gas chamber was flushed with $N_2$ and the 1σ standard deviation of the 2f PTS signal was measured for 6 setpoints of the *pump* power with 1 s averaging. The results are plotted in Fig. 7a. The 2f signal amplitude increases approximately linearly with the *pump* power, which is consistent with Equation 1. The amplitude of the recorded noise (measured in $N_2$) was between 31 μV and 55 μV at all set points. This confirms that the use of the maximum output power available from the BOA (66 mW) does not deteriorate the SNR of the sensor and proves that the pump does not clip the mirrors of the FFPC. Furthermore, by performing full 2f scans, we confirmed that the developed sensor is capable of registering the PTS signal also with *pump* sources delivering low output power (signal registered with 1.7 mW excitation power is depicted in the inset in Fig. 7b). Note, that the measurement was performed with 50 ms time constant (TC) set on the LIA, thus random noise is visible. After the optimal measurement parameters were determined, the performance of the constructed sensor was evaluated.

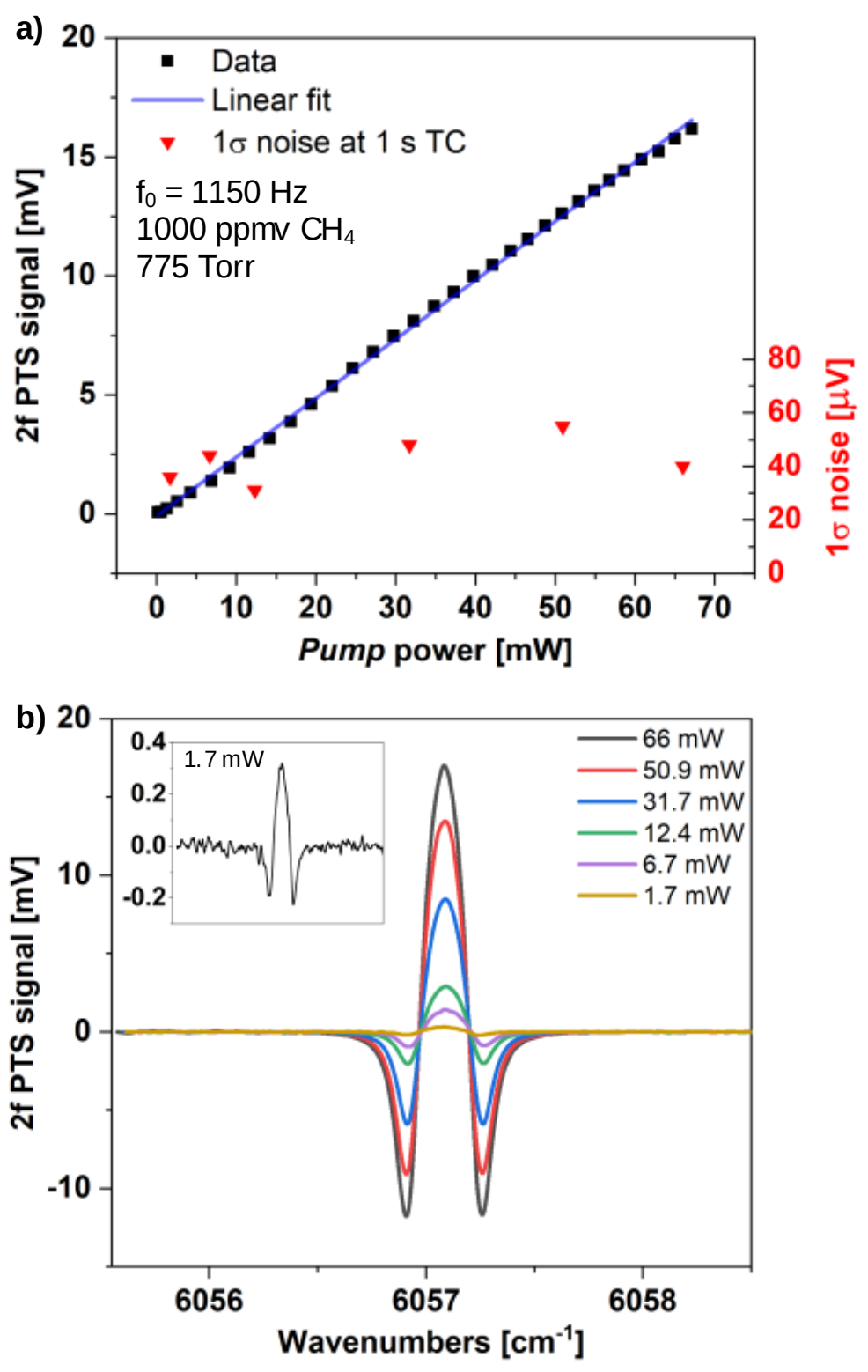

Figure 7. (a) Maximum 2f PTS signal amplitude measured as a function of *pump* power used to excite $CH_4$ molecules inside the FFPC. 1σ standard deviation of the signal measured after flushing the gas chamber with $N_2$ is plotted with red triangles. The points in the graph were averaged for 1 s. (b) Full 2f spectra of the targeted absorption line registered for several setpoints of the *pump* power. Crucial measurement details are listed on the graphs.

### 3.3 Sensor performance evaluation

In order to calculate the minimum detection limit (MDL) of the sensor, its response for a certified gas mixture has to be measured and compared with the sensor noise. For this measurement the parameters optimized in Section 3.2 were used to record a full 2f spectrum for a mixture of 1000 ppmv of $CH_4$ in $N_2$ at 775 Torr. Additionally, the flatness of the sensor baseline was assessed by repeating the measurement with pure $N_2$ flowing through the gas chamber. During the measurements, a sampling rate of 1.764 kHz was used along with a time constant (TC) setting of 100 ms. The results are plotted in Fig. 8.

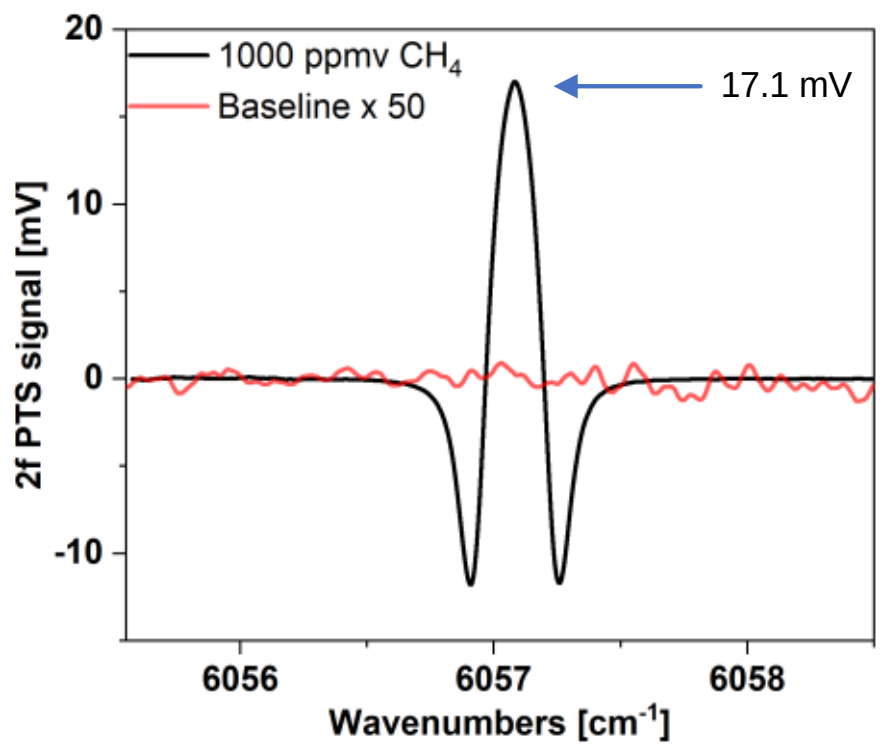

Figure 8. The black plot shows a full 2f WMS spectrum of the $CH_4$ absorption line located at 6057.1 $cm^{-1}$ registered using optimized measurement parameters (listed in the graph). The LIA TC was set to 100 ms. The red trace shows the sensor baseline measured with $N_2$ flowing through the gas chamber (magnified x50 for visibility).

A maximum 2f PTS signal amplitude of 17.1 mV was obtained for a *pump* power of 66 mW for a mixture of 1000 ppmv $CH_4$ in $N_2$ at a pressure of 775 Torr. The sensor baseline does not show fringe indications. In the second step, the linearity of the sensor response was measured by preparing 7 gas samples containing $CH_4$ diluted in $N_2$, using the gas mixer. During the measurement, the *pump* laser was tuned to the center of the $CH_4$ transition located at ~6057.1 $cm^{-1}$ and each plotted measurement point was averaged for 60 s. The results are plotted in Fig. 9.

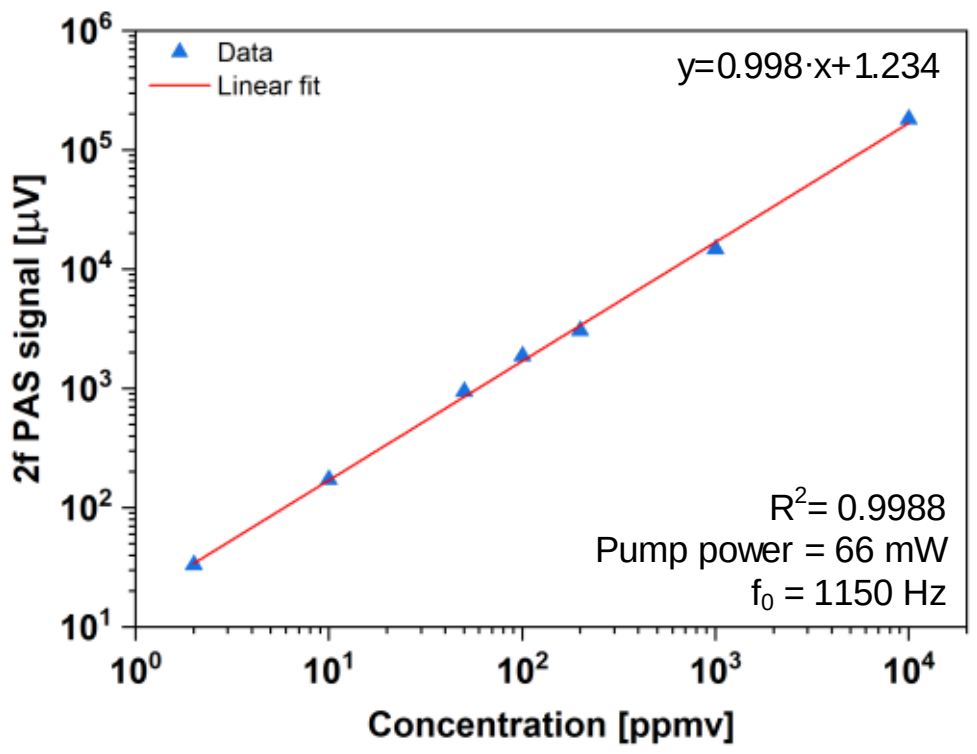

Figure 9. Maximum amplitude of registered 2f PTS signals plotted as a function of $CH_4$ concentration flowing through the gas chamber. The solid line represents a linear fit of the measured values. The measurement parameters are listed in the graph.

The measurement confirmed that the amplitude of the registered 2f PTS signal increases approximately linearly with the measured gas concentration, resulting in an $R^2$ value of a linear fit of 0.9988. We believe that the gas mixer used to prepare the samples is the main source of the deviations observed on the graph. However, the obtained linearity is comparable to previously published experimental work on PTS [39,40]. The last experiment involved recording the long-term sensor noise. The gas chamber was flushed with $N_2$ and a constant flow of 100 sccm was maintained. The *pump* laser was tuned to the center of the $CH_4$ transition and the 2f PTS signal was measured for 30 minutes with 1.032 ms TC and 1.764 kHz sampling set on the LIA. The noise floor of the reflection detector and the UHFLI-LIA was registered independently. The Allan deviation was calculated based on the measurements and is plotted in Fig. 10.

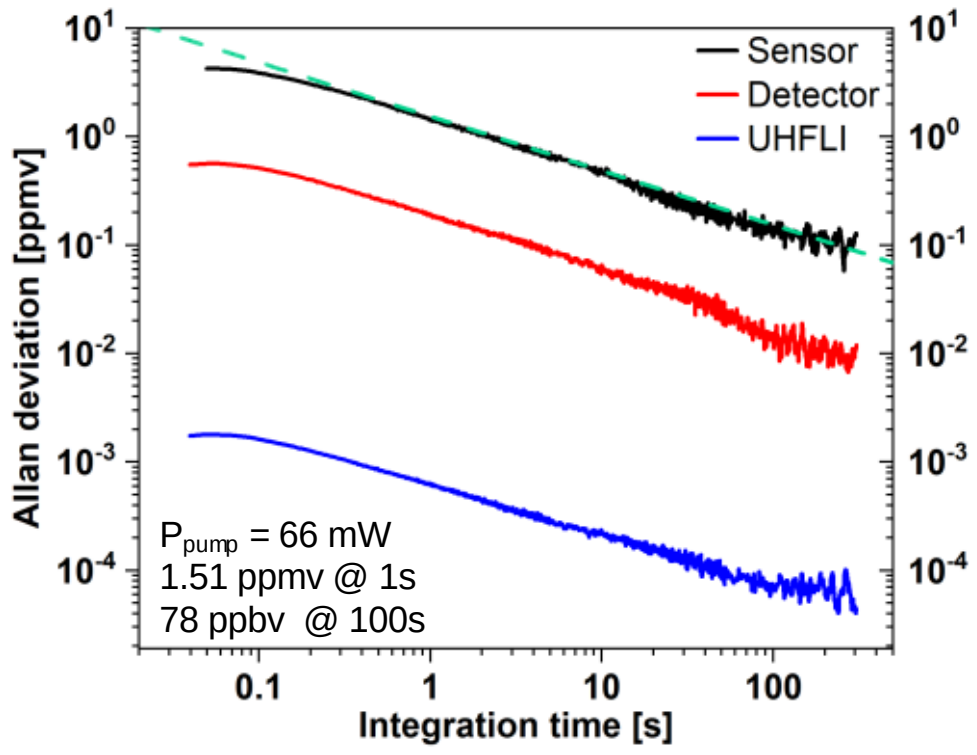

Figure 10. Allan deviation plots. The graphs were calculated separately based on the sensor noise in $N_2$, noise floor of the detector measuring the reflection from the FFPC and UHFLI noise (input terminated with 50 Ohm balun). The Y axis was calculated in ppmv for clarity, based on the results from the plot in Fig. 8. The dashed green line shows the $1/\sqrt{\tau}$ trend.

Based on the calculated Allan deviation, the MDL of the sensor reached 1.51 ppmv and 78 ppbv of $CH_4$ for 1 s and 100 s integration time, respectively. This corresponds to a noise equivalent absorption coefficient (NEA) of $5.54\times10^{-7}$ $cm^{-1}$ and $2.99\times10^{-8}$ $cm^{-1}$, for the integration time of 1 s and 100 s, respectively. Note that the Allan deviation characteristic of the sensor does not significantly deviate from the $1/\sqrt{\tau}$ approximation, even for integration times reaching 300 s Similar Allan deviation trends were recorded for all components of the setup (sensor, detector, UHFLI) as shown in Fig. 10. As this indicates uncorrelated noise sources, longer signal averaging can be utilized to reach lower MDL. For gas sensors using the PTS and PAS techniques, the normalized noise equivalent absorption (NNEA) parameter is calculated, which standardizes the detectivity of the device based on the gas absorption line strength and the optical power of the *pump* laser used to induce the RI modulation [35,41]:

$$NNEA = \frac{\alpha_{min} P_{exc}}{\sqrt{ENBW}} \tag{4}$$

where $\alpha_{min}$ is the minimum detectable absorption coefficient, which was calculated using the HITRAN database ($2.99\times10^{-8}$ $cm^{-1}$ @ 100 s integration time) and ENBW is the equivalent noise bandwidth, which depends on the filtering settings used for signal acquisition (0.78 mHz for TC = 100 s and 24 dB/oct). The NNEA for $P_{exc}$ = 66 mW reached $7.06\times10^{-8}$ $cm^{-1}$ W $Hz^{-1/2}$. An additional experiment was performed to estimate the value of the RI modulation induced in the gas sample. For low gas concentrations the modulation value is small and requires LIA-based signal extraction, therefore for this experiment pure $CH_4$ was flowing through the gas chamber and 2.39 mW of *pump* power was used to excite the molecules (taking into account the 64.9% transmission of the gas over the path length between the wedged window and the FFPC inside the gas chamber). Initially, the FFPC was scanned using a 4 Hz triangle voltage signal applied to its PZT element, while the *pump* wavelength was modulated with a sine wave signal ($f_0$ = 1150 Hz) around the $CH_4$ absorption peak (with a 10.7 GHz modulation depth) – as shown in Figure 11a. This allowed us to record the complete FFPC reflection with the PT-induced RI modulation clearly visible as sinusoidal variations in the reflection signal (inset in Figure 11a). The effect of RI modulation on the FFPC reflection signal amplitude is also illustrated in the graph for clarity. To determine the value of the induced RI modulation, it was necessary to correlate it with the observed variations in the reflection signal. This was achieved by simulating the FFPC reflection signal as a function of the RI change, based on mirror separation and finesse, according to [42]. The simulation was then fitted to the measured cavity reflectance (red plot in Figure 11a), providing information on the RI modulation induced in the pure $CH_4$ gas sample. In the next experiment, two measurements were performed in which the FFPC was successively locked on both slopes of the cavity reflection. The locking positions that result in the maximum modulation of the reflection signal were chosen, as described in Section 3.1 and are highlighted as A and B points in Figure 11a. Subsequently, the *pump* laser was modulated with a square-shaped signal at f = 1150 Hz, resulting in modulation of the reflection signal, as depicted in Figure 11b. The difference in the observed amplitudes results from the dissimilar steepness of the FFPC reflection signal, which is evident from the measurement shown in Figure 3. For pure $CH_4$ and 2.39 mW *pump* power exciting the molecules, an RI modulation of $\Delta n = 8.21\times10^{-7}$ was observed when the sensor was locked to the 'B' point on the dip. According to Equation 1, the PT signal scales linearly with the *pump* power and the gas absorption coefficient. Therefore, we can estimate the RI modulation sensitivity of the constructed sensor at values of $3.42\times10^{-11}$ and $1.77\times10^{-12}$ for integration time of 1 s and 100 s, respectively. The obtained values are comparable to our previous micro-PTS gas sensor, which had a significantly more complex layout and required sophisticated signal extraction [43].

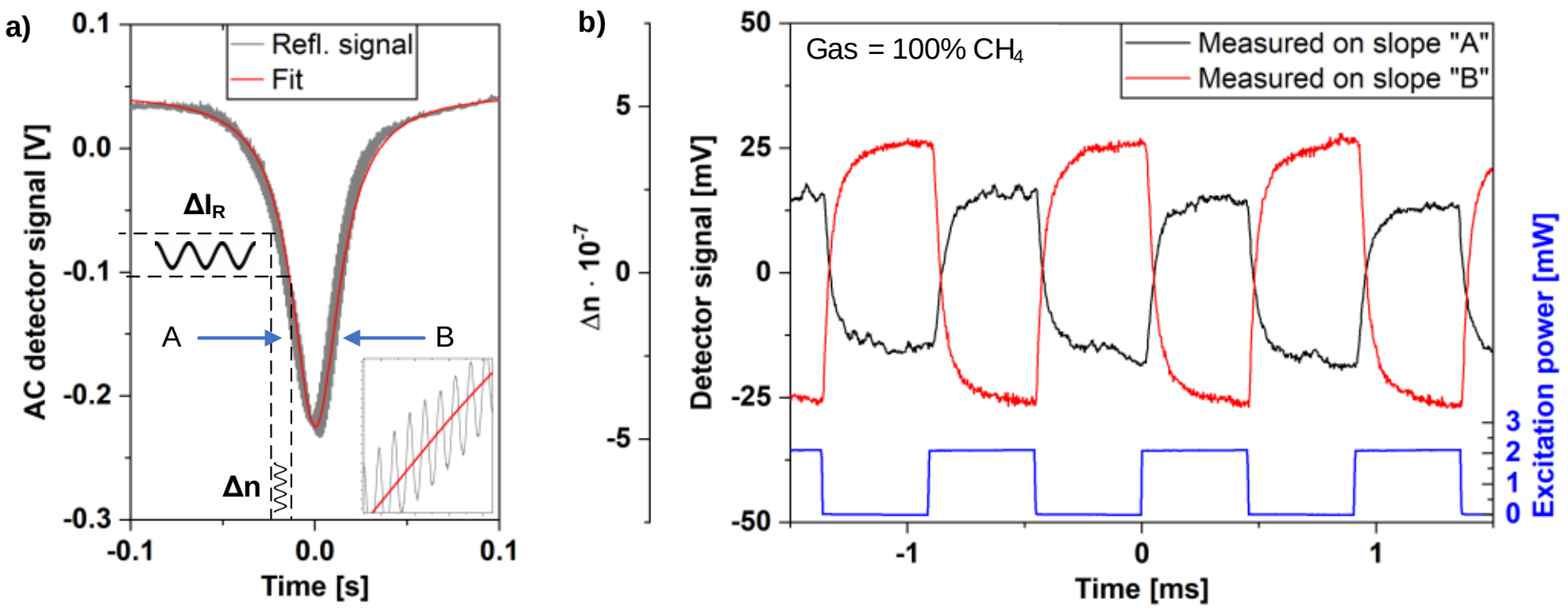

Figure 11. (a) – full FFPC reflection signal registered with a 4 Hz sawtooth ramp voltage applied to the PZT stretching element. The signal is fitted with a Lorentzian function to estimate the induced RI modulation in the gas sample. The inset magnifies a section of the "B" slope, highlighting the effect of the RI modulation on the amplitude of the cavity reflection signal. (b) – modulation of the $I_R$ measured at the optimal locking points on both slopes of the FFPC reflection. The PT effect was induced by modulating the *pump* laser with a square-shaped signal at a frequency of 1150 Hz. During the measurements pure $CH_4$ was flowing through the cell and 2.39 mW of *pump* power was used.

## 4. Discussion and conclusions

We described and experimentally verified the first demonstration of implementing a miniaturized fiber cavity for precise photothermal signal extraction. We enclosed a simple, self-developed, robust and stable FFPC designed to work at wavelengths ~780 nm in a custom-designed aluminum gas chamber, which provided repeatable measurement conditions. In the proof-of-concept experiment, we used an auxiliary near-IR pump laser to excite $CH_4$ molecules at their transition located at 6057.1 cm$^{-1}$. The emission of an off-the-shelf DFB source was boosted in a BOA to a value of 66 mW. We implemented the WMS technique to simplify signal extraction and simultaneously improve SNR. The *pump* laser wavelength was modulated with a sinewave signal ($f_0$) and the amplitude of the RI modulation was encoded onto the FFPC reflection signal monitored by a standard Si photodetector. The LIA demodulated the signal at the second harmonic of the *pump* modulation ($2\times f_0$), similarly to classical WMS configurations. The PTS technique is a non-direct detection method, thus appropriate design of the sensor should result in a flat baseline. This was true for the presented sensor and was confirmed twofold, by registering a full 2f sweep with $N_2$ flowing through the gas chamber (Fig. 8) and by calculating the Allan deviation (Fig. 10). As a result, we achieved sub-ppmv detection capabilities, which are comparable to previously published sensors that relied on the PT effect. Worth pointing out is the fact that the sensors principle of operation depends on probing the PT-induced RI modulation in a miniscule region, which overlaps with the standing wave mode-profile of the FFPC, which has a waist radius of only ~ 4.2 μm in the center of the 106 μm long-interferometer. This makes the design significantly smaller and less complex, compared to our previous miniaturized PTS gas sensor, which required building a complete solid-state laser with two separate laser emissions in the cavity [43]. Moreover, the current version of the sensor does not require additional RF components, like mixers, external local oscillators and amplifiers as the spectroscopic signal is directly demodulated from the photocurrent of an off-the-shelf detector. We optimized the sensor parameters in a series of experiments, resulting in an MDL of $CH_4$ at the level of 78 ppbv, which is equal to an NEA of $2.99\times10^{-8}$ cm$^{-1}$ at 100 s integration time. For 100 s integration time, the NNEA reached $7.06\times10^{-8}$ cm$^{-1}$ W Hz$^{-1/2}$. The obtained detectivity of the sensor is comparable to that of other sensors that utilize the PT effect for gas detection. By simulating and fitting a curve to the measured reflection signal we estimated the RI modulation detection limit of the constructed sensor, which reached $1.77\times10^{-12}$ for an integration time of 100 s. Table 1 compares the results with previously published near-IR sensors relying on the PTS effect and a MZI or FPI signal readout.

**Table 1. Comparison of results on near-IR sensors relying on the PTS effect and a MZI or FPI signal readout**

| **Configuration** | **Gas** | **λ [nm]** | **NEA [cm$^{-1}$]** | **Ref.** |
|---|---|---|---|---|
| Hollow-core fiber Mach-Zehnder interferometer | Acetylene | 1530.37 | $2.3 \times 10^{-9}$ | [4] |
| Antiresonant Hollow-core fiber Mach-Zehnder interferometer | Carbon monoxide | 2327 | $3.2 \times 10^{-6}$ | [5] |
| Hollow-core fiber Fabry-Perot interferometer | Acetylene | 1531.58 | $7.43\times 10^{-9}$ | [9] |
| Hollow-core fiber Fabry-Perot interferometer | Methane | 1650.65 | $1.35 \times 10^{-8}$ | [9] |
| Antiresonant Hollow-core fiber Fabry-Perot interferometer | Acetylene | 1530.37 | $2.3 \times 10^{-9}$ | [10] |
| Hollow-core fiber Fabry-Perot interferometer | Acetylene | 1530.37 | $1.26 \times 10^{-7}$ | [11] |
| Hollow-core fiber Fabry-Perot interferometer | Acetylene | 1530.37 | $2.7 \times 10^{-7}$ | [12] |
| Fiber Fabry-Perot Interferometer | Methane | 1651 | $2.99\times10^{-8}$ | This paper |

Our proposed sensor configuration demonstrated performance comparable to previously published PTS sensors using MZI or FPI spectroscopic signal extraction methods. Nevertheless, the competitors' achievement of a low NEA primarily stems from their utilization of all-in-fiber sensor design, consequently restricting their operational wavelength range to the near-IR region, thus impeding their versatility. Based on the preliminary results, we believe that the detection capability of the new sensor design can be significantly enhanced. First, a high finesse FFPC will be tested in the sensor. Finesse values greater than $10^5$ were obtained in similar cavity configurations, which have reflection curves with steeper slopes, compared to the cavity used in this experiment [16,17]. This will increase the $\Delta I_R/\Delta n$ ratio, which is directly correlated with the amplitude of the registered spectroscopic signal. Moreover, higher finesse results in narrower transmission peaks of the cavities (~9 MHz, compared to ~1 GHz in this paper), thus simple phase modulation of the *probe* beam (e.g. with a fiberized modulator) will allow the use of the Pound-Drever-Hall (PDH) cavity locking technique, as presented in [18]. The steep error signal centered on the cavity transmission peak should improve the PID lock, thus enhance the overall long-term stability of the sensor. Lastly, due to the non-direct detection of the spectroscopic signal, the sensor has no limitations in terms of the *pump* laser wavelength if the design is properly adjusted. This allows measurement of any gas, provided an appropriate excitation source is available to induce the PT effect inside the air gap of the FFPC, making the sensor a versatile platform for detecting molecules in the wavelength range from UV to THz. With some modifications, the proposed sensor configuration would be capable of measuring gas concentrations in practical applications, e.g. ambient methane. A small gas pump could be added to the system to force the atmosphere air into the gas chamber, or the gas chamber could be removed, to expose the sensor to the environment. Future work will focus on demonstrating long-term measurements of gases present in ambient air.

Experiments with excitation lasers emitting in the mid-IR wavelength region, abundant in strong absorption lines of numerous species, will be the focus of our subsequent experiments.

## Data availability

Data supporting the findings of this study are available from the corresponding author upon reasonable request.

## CRediT authorship contribution statement

Karol Krzempek: Conceptualization, Data curation, Formal analysis, Investigation, Methodology, Project administration, Resources, Software, Supervision, Validation, Visualization, Writing - original draft, Writing - review & editing. Piotr Jaworski: Funding acquisition, Resources, Writing - review & editing. Lukas Tenbrake: fiber surface machining, FFPC manufacturing, FFPC characterization, writing – review & editing. Florian Giefer: fiber surface machining, FFPC manufacturing, FFPC characterization, writing – review & editing. Dieter Meschede: funding acquisition, FFPC device design, writing – review & editing. Sebastian Hofferberth: funding acquisition, writing – review & editing. Hannes Pfeifer: FFPC device design, fiber surface machining, funding acquisition, writing – review & editing.

**Declaration of Competing Interest**

The authors report no declarations of interest.

**Acknowledgements**

This work was supported by the National Science Centre, Poland under "M-ERA.NET 2 Call 2019" (grant number 2019/01/Y/ST7/00088) and grant number 2019/35/D/ST7/04436. Deutsche Forschungsgemeinschaft (DFG, German Research Foundation) under Germany's Excellence Strategy – Cluster of Excellence Matter and Light for Quantum Computing (ML4Q) EXC 2004/1 – 390534769; BMBF, Federal Ministry for Education and Research - project FaResQ; European Union's Horizon 2020 ERC consolidator grant - RYD-QNLO - Grant No. 771417.

**References**

[1] P. Saxena, P. Shukla, A Review on Gas Sensor Technology and Its Applications, in: V.V. Rao, A. Kumaraswamy, S. Kalra, A. Saxena (Eds.), Computational and Experimental Methods in Mechanical Engineering, Springer, Singapore, 2022: pp. 165–175. https://doi.org/10.1007/978-981-16-2857-3_18.

[2] W. Demtröder, Laser Spectroscopy: Vol. 1: Basic Principles, 4th ed., Springer-Verlag, Berlin Heidelberg, 2008. https://doi.org/10.1007/978-3-540-73418-5.

[3] K. Krzempek, A review of photothermal detection techniques for gas sensing applications, Applied Sciences. 9 (2019) 2826.

[4] W. Jin, Y. Cao, F. Yang, H.L. Ho, Ultra-sensitive all-fibre photothermal spectroscopy with large dynamic range, Nature Communications. 6 (2015) 6767. https://doi.org/10.1038/ncomms7767.

[5] C. Yao, Q. Wang, Y. Lin, W. Jin, L. Xiao, S. Gao, Y. Wang, P. Wang, W. Ren, Photothermal CO detection in a hollow-core negative curvature fiber, Opt. Lett., OL. 44 (2019) 4048–4051. https://doi.org/10.1364/OL.44.004048.

[6] C. Yao, S. Gao, Y. Wang, P. Wang, W. Jin, W. Ren, MIR-Pump NIR-Probe Fiber-Optic Photothermal Spectroscopy With Background-Free First Harmonic Detection, IEEE Sensors Journal. 20 (2020) 12709–12715. https://doi.org/10.1109/JSEN.2020.3000795.

[7] Z. Li, Z. Wang, F. Yang, W. Jin, W. Ren, Mid-infrared fiber-optic photothermal interferometry, Opt. Lett., OL. 42 (2017) 3718–3721. https://doi.org/10.1364/OL.42.003718.

[8] K. Krzempek, Part-per-billion level photothermal nitric oxide detection at 5.26 µm using antiresonant hollow-core fiber-based heterodyne interferometry, Optics Express. 29 (2021) 32568–32579.

[9] F. Liu, H. Bao, H.L. Ho, W. Jin, S. Gao, Y. Wang, Multicomponent trace gas detection with hollow-core fiber photothermal interferometry and time-division multiplexing, Opt. Express, OE. 29 (2021) 43445–43453. https://doi.org/10.1364/OE.446478.

[10] H. Bao, H. Bao, Y. Hong, Y. Hong, W. Jin, W. Jin, H.L. Ho, H.L. Ho, C. Wang, C. Wang, S. Gao, Y. Wang, P. Wang, Modeling and performance evaluation of in-line Fabry-Perot photothermal gas sensors with hollow-core optical fibers, Opt. Express, OE. 28 (2020) 5423–5435. https://doi.org/10.1364/OE.385670.

[11] Y. Tan, W. Jin, F. Yang, Y. Jiang, H.L. Ho, Cavity-Enhanced Photothermal Gas Detection With a Hollow Fiber Fabry-Perot Absorption Cell, Journal of Lightwave Technology. 37 (2019) 4222–4228. https://doi.org/10.1109/JLT.2019.2922001.

[12] F. Yang, Y. Tan, W. Jin, Y. Lin, Y. Qi, H.L. Ho, Hollow-core fiber Fabry–Perot photothermal gas sensor, Opt. Lett., OL. 41 (2016) 3025–3028. https://doi.org/10.1364/OL.41.003025.

[13] Z. Yang, G. Yin, C. Liang, T. Zhu, Photothermal Interferometry Gas Sensor Based on the First Harmonic Signal, IEEE Photonics Journal. 10 (2018) 1–7. https://doi.org/10.1109/JPHOT.2018.2821845.

[14] K. Krzempek, P. Jaworski, P. Kozioł, W. Belardi, Antiresonant hollow core fiber-assisted photothermal spectroscopy of nitric oxide at 5.26 µm with parts-per-billion sensitivity, Sensors and Actuators B: Chemical. 345 (2021) 130374. https://doi.org/10.1016/j.snb.2021.130374.

[15] F. Chen, F. Chen, S. Jiang, W. Jin, W. Jin, H. Bao, H. Bao, H.L. Ho, H.L. Ho, C. Wang, C. Wang, S. Gao, Ethane detection with mid-infrared hollow-core fiber photothermal spectroscopy, Opt. Express, OE. 28 (2020) 38115–38126. https://doi.org/10.1364/OE.410927.

[16] H. Pfeifer, L. Ratschbacher, J. Gallego, C. Saavedra, A. Faßbender, A. von Haaren, W. Alt, S. Hofferberth, M. Köhl, S. Linden, D. Meschede, Achievements and perspectives of optical fiber Fabry–Perot cavities, Appl. Phys. B. 128 (2022) 29. https://doi.org/10.1007/s00340-022-07752-8.

[17] C. Saavedra, D. Pandey, W. Alt, H. Pfeifer, D. Meschede, Tunable fiber Fabry-Perot cavities with high passive stability, Opt. Express, OE. 29 (2021) 974–982. https://doi.org/10.1364/OE.412273.

[18] C. Saavedra, D. Pandey, W. Alt, D. Meschede, H. Pfeifer, Spectroscopic Gas Sensor Based on a Fiber Fabry-Perot Cavity, Phys. Rev. Appl. 18 (2022) 044039. https://doi.org/10.1103/PhysRevApplied.18.044039.

[19] E.R. Abraham, E.A. Cornell, Teflon feedthrough for coupling optical fibers into ultrahigh vacuum systems, Appl Opt. 37 (1998) 1762–1763. https://doi.org/10.1364/ao.37.001762.

[20] J.M. Supplee, E.A. Whittaker, W. Lenth, Theoretical description of frequency modulation and wavelength modulation spectroscopy, Appl. Opt., AO. 33 (1994) 6294–6302. https://doi.org/10.1364/AO.33.006294.

[21] W.H. Flygare, Molecular relaxation, Acc. Chem. Res. 1 (1968) 121–127. https://doi.org/10.1021/ar50004a004.

[22] A. Zifarelli, R. De Palo, P. Patimisco, M. Giglio, A. Sampaolo, S. Blaser, J. Butet, O. Landry, A. Müller, V. Spagnolo, Multi-gas quartz-enhanced photoacoustic sensor for environmental monitoring exploiting a Vernier effect-based quantum cascade laser, Photoacoustics. 28 (2022) 100401. https://doi.org/10.1016/j.pacs.2022.100401.

[23] G. Wu, Z. Gong, J. Ma, H. Li, M. Guo, K. Chen, W. Peng, Q. Yu, L. Mei, High-sensitivity miniature dual-resonance photoacoustic sensor based on silicon cantilever beam for trace gas sensing, Photoacoustics. 27 (2022) 100386. https://doi.org/10.1016/j.pacs.2022.100386.

[24] Z. Wang, Q. Wang, H. Zhang, S. Borri, I. Galli, A. Sampaolo, P. Patimisco, V.L. Spagnolo, P. De Natale, W. Ren, Doubly resonant sub-ppt photoacoustic gas detection with eight decades dynamic range, Photoacoustics. 27 (2022) 100387. https://doi.org/10.1016/j.pacs.2022.100387.

[25] S. Qiao, A. Sampaolo, P. Patimisco, V. Spagnolo, Y. Ma, Ultra-highly sensitive HCl-LITES sensor based on a low-frequency quartz tuning fork and a fiber-coupled multi-pass cell, Photoacoustics. 27 (2022) 100381. https://doi.org/10.1016/j.pacs.2022.100381.

[26] S. Li, J. Lu, Z. Shang, X. Zeng, Y. Yuan, H. Wu, Y. Pan, A. Sampaolo, P. Patimisco, V. Spagnolo, L. Dong, Compact quartz-enhanced photoacoustic sensor for ppb-level ambient NO2 detection by use of a high-power laser diode and a grooved tuning fork, Photoacoustics. 25 (2022) 100325. https://doi.org/10.1016/j.pacs.2021.100325.

[27] L. Fu, P. Lu, C. Sima, J. Zhao, Y. Pan, T. Li, X. Zhang, D. Liu, Small-volume highly-sensitive all-optical gas sensor using non-resonant photoacoustic spectroscopy with dual silicon cantilever optical microphones, Photoacoustics. 27 (2022) 100382. https://doi.org/10.1016/j.pacs.2022.100382.

[28] L. Liu, H. Huan, W. Li, A. Mandelis, Y. Wang, L. Zhang, X. Zhang, X. Yin, Y. Wu, X. Shao, Highly sensitive broadband differential infrared photoacoustic spectroscopy with wavelet denoising algorithm for trace gas detection, Photoacoustics. 21 (2021) 100228. https://doi.org/10.1016/j.pacs.2020.100228.

[29] Z. Gong, T. Gao, L. Mei, K. Chen, Y. Chen, B. Zhang, W. Peng, Q. Yu, Ppb-level detection of methane based on an optimized T-type photoacoustic cell and a NIR diode laser, Photoacoustics. 21 (2021) 100216. https://doi.org/10.1016/j.pacs.2020.100216.

[30] X. Zhao, C. Li, H. Qi, J. Huang, Y. Xu, Z. Wang, X. Han, M. Guo, K. Chen, Integrated near-infrared fiber-optic photoacoustic sensing demodulator for ultra-high sensitivity gas detection, Photoacoustics. 33 (2023) 100560. https://doi.org/10.1016/j.pacs.2023.100560.

[31] K. Chen, Q. Yu, Z. Gong, M. Guo, C. Qu, Ultra-high sensitive fiber-optic Fabry-Perot cantilever enhanced resonant photoacoustic spectroscopy, Sensors and Actuators B: Chemical. 268 (2018) 205–209. https://doi.org/10.1016/j.snb.2018.04.123.

[32] D. Pinto, H. Moser, J.P. Waclawek, S. Dello Russo, P. Patimisco, V. Spagnolo, B. Lendl, Parts-per-billion detection of carbon monoxide: A comparison between quartz-enhanced photoacoustic and photothermal spectroscopy, Photoacoustics. 22 (2021) 100244. https://doi.org/10.1016/j.pacs.2021.100244.

[33] A.J. Campillo, S.J. Petuchowski, C.C. Davis, H. Lin, Fabry–Perot photothermal trace detection, Applied Physics Letters. 41 (1982) 327–329. https://doi.org/10.1063/1.93524.

[34] J. Gallego, S. Ghosh, S.K. Alavi, W. Alt, M. Martinez-Dorantes, D. Meschede, L. Ratschbacher, High-finesse fiber Fabry–Perot cavities: stabilization and mode matching analysis, Appl. Phys. B. 122 (2016) 47. https://doi.org/10.1007/s00340-015-6281-z.

[35] J.P. Waclawek, H. Moser, B. Lendl, Balanced-detection interferometric cavity-assisted photothermal spectroscopy employing an all-fiber-coupled probe laser configuration, Opt. Express, OE. 29 (2021) 7794–7808. https://doi.org/10.1364/OE.416536.
[36] L.S. Rothman, I.E. Gordon, A. Barbe, D.C. Benner, P.F. Bernath, M. Birk, V. Boudon, L.R. Brown, A. Campargue, J.-P. Champion, K. Chance, L.H. Coudert, V. Dana, V.M. Devi, S. Fally, J.-M. Flaud, R.R. Gamache, A. Goldman, D. Jacquemart, I. Kleiner, N. Lacome, W.J. Lafferty, J.-Y. Mandin, S.T. Massie, S.N. Mikhailenko, C.E. Miller, N. Moazzen-Ahmadi, O.V. Naumenko, A.V. Nikitin, J. Orphal, V.I. Perevalov, A. Perrin, A. Predoi-Cross, C.P. Rinsland, M. Rotger, M. Šimečková, M.A.H. Smith, K. Sung, S.A. Tashkun, J. Tennyson, R.A. Toth, A.C. Vandaele, J. Vander Auwera, The HITRAN 2008 molecular spectroscopic database, Journal of Quantitative Spectroscopy and Radiative Transfer. 110 (2009) 533–572. https://doi.org/10.1016/j.jqsrt.2009.02.013.
[37] P. Bojęś, P. Pokryszka, P. Jaworski, F. Yu, D. Wu, K. Krzempek, Quartz-Enhanced Photothermal Spectroscopy-Based Methane Detection in an Anti-Resonant Hollow-Core Fiber, Sensors. 22 (2022) 5504.
[38] G.B. Rieker, J.B. Jeffries, R.K. Hanson, Calibration-free wavelength-modulation spectroscopy for measurements of gas temperature and concentration in harsh environments, Appl. Opt., AO. 48 (2009) 5546–5560. https://doi.org/10.1364/AO.48.005546.
[39] C. Yao, S. Gao, Y. Wang, W. Jin, W. Ren, Heterodyne interferometric photothermal spectroscopy for gas detection in a hollow-core fiber, Sensors and Actuators B: Chemical. 346 (2021) 130528. https://doi.org/10.1016/j.snb.2021.130528.
[40] M. Hu, C. Yao, A. Ventura, J.G. Hayashi, F. Poletti, W. Ren, Mid-Infrared Photothermal Gas Sensor Enabled By Core-Cladding Mode Interference in a Hollow-Core Fiber, J. Lightwave Technol., JLT. 40 (2022) 6568–6575.
[41] K. Krzempek, A Review of Photothermal Detection Techniques for Gas Sensing Applications, Applied Sciences. 9 (2019) 2826. https://doi.org/10.3390/app9142826.
[42] J.P. Waclawek, V.C. Bauer, H. Moser, B. Lendl, 2f-wavelength modulation Fabry-Perot photothermal interferometry, Opt. Express, OE. 24 (2016) 28958–28967. https://doi.org/10.1364/OE.24.028958.
[43] G. Dudzik, K. Krzempek, K. Abramski, G. Wysocki, Solid-state laser intra-cavity photothermal gas sensor, Sensors and Actuators B: Chemical. 328 (2021) 129072. https://doi.org/10.1016/j.snb.2020.129072.